\documentclass[pdflatex,sn-mathphys-ay]{sn-jnl}

\usepackage{graphicx}%
\usepackage{multirow}%
\usepackage{amsmath,amssymb,amsfonts}%
\usepackage{amsthm}%
\usepackage{mathrsfs}%
\usepackage[title]{appendix}%
\usepackage{xcolor}%
\usepackage{textcomp}%
\usepackage{manyfoot}%
\usepackage{booktabs}%
\usepackage{algorithm}%
\usepackage{algorithmicx}%
\usepackage{algpseudocode}%
\usepackage{listings}
\usepackage{csquotes}%
\usepackage{makeidx}

\theoremstyle{thmstyleone}%
\theoremstyle{thmstyletwo}%

\theoremstyle{thmstylethree}%

\begin{document}

\noindent\fbox{%
  \parbox{\textwidth}{%
    \textbf{Note.} This is the longer version of a chapter that will appear in the edited volume \textit{How to Understand Quantum Mechanics – 100 Years of Ongoing Interpretation}, edited by Lars–Göran Johansson and Jan Faye for the \textit{Boston Studies in the Philosophy and History of Science} series. Please only cite the published version.
  }%
}

\title[The Mathematization of Nature]{Wave Function Realism and the Mathematization of Nature. A Phenomenological Perspective}


\author*[1]{\fnm{Harald A.} \sur{Wiltsche}}\email{harald.wiltsche@liu.se}

\author[2]{\fnm{Philipp} \sur{Berghofer}}\email{philipp.berghofer@uni-graz.at}
\equalcont{These authors contributed equally to this work.}

\affil*[1]{\orgdiv{Division of Philosophy and Applied Ethics}, \orgname{Linköping University}, \orgaddress{\street{Campus Valla}, \city{Linköping}, \postcode{581 83}, \country{Sweden}}}

\affil[2]{\orgdiv{Department of Philosophy}, \orgname{University of Graz}, \orgaddress{\street{Heinrichstraße 26/5}, \city{Graz}, \postcode{8010}, \country{Austria}}}


\abstract{This chapter reexamines wave function realism (WFR) through the lens of phenomenology\index{phenomenology}. We begin by situating WFR within the broader debate about the ontology of the quantum state and the temptation to \enquote{read off} metaphysics from mathematical formalism. Against this background, we turn to the London–Bauer interpretation (LBI\index{London-Bauer-Interpretation}), the most explicit attempt to interpret quantum mechanics through phenomenological categories. On this view, the measurement transition is not a physical discontinuity but a reflective articulation of objectivity, and the wave function\index{wave function} formally encodes the horizonal structure of world-givenness. We develop this idea by reconfiguring the notion of realism itself: not as objectivist, but as correlational and transcendental\index{transcendental}. The resulting picture suggests that quantum mechanics, rather than depicting a world \enquote{minus observers,} mathematically articulates the very correlation through which a world becomes manifest at all.}


\maketitle

\section{Introduction}\label{sec1}

It has become something of a tradition in the foundations of quantum mechanics to begin by noting that, for all its extraordinary empirical success, quantum theory continues to resist consensus on what, if anything, it tells us about the world. In fact, there is good reason for the gesture: it \textit{is} genuinely puzzling that a theory of such predictive and technological power should prove so obstinately opaque when it comes to interpretation. Apart from adding to the chorus of those who point out the obvious, our aim in this chapter is to contribute to a discussion that has been central in recent decades and is, in many ways, symptomatic of deeper tensions within the philosophy of quantum mechanics: the issue of wave function realism \index{wave function realism} (hereafter \textit{WFR}). That this particular debate continues to attract such sustained attention suggests not only the difficulty of interpreting quantum theory, but also a persistent unease about what kind of metaphysical commitments are—explicitly or implicitly—licensed by the formalism.

The debate about WFR is well established in the field, and we will return to some aspects of it later. What distinguishes our contribution, however, is the metaphilosophical framework from which our analysis will proceed. As the title of our chapter suggests, we approach WFR from a specifically \textit{phenomenological} angle—a choice that, given the somewhat underrepresented status of phenomenology\index{phenomenology} in contemporary philosophy of physics, may seem unexpected. Still, bringing phenomenology\index{phenomenology} into the mix of available metaphilosophical frameworks in the philosophy of quantum mechanics seems worthwhile for at least two reasons. First, as will become clearer in what follows, phenomenology\index{phenomenology} has in fact played a role in the history of quantum interpretations—most notably through the work of Fritz London\index{London, Fritz} and Edmond Bauer. More recently, there have also been attempts to draw on genuinely phenomenological resources to elucidate programs such as QBism\index{QBism} \citep[]{BerghoferWiltsche2023} and quantum reconstruction \citep[]{BerghoferGoyalWiltsche2020}. The fact that these developments remain relatively unknown in mainstream circles is largely due to contingent factors—such as historically entrenched allegiances to labels like \enquote{analytic} or \enquote{continental} philosophy. For this reason, engaging with the phenomenological tradition seems relevant to appreciate the full breadth of the debate in quantum foundations. Second, our central philosophical claim in this chapter is that phenomenology\index{phenomenology} has the potential to significantly expand the range of theoretical options in the debate about WFR.

The structure of our chapter is as follows: In \hyperref[sec2]{Section 2}, we briefly outline key aspects of the WFR debate, including the standard range of theoretical options and the reasons why many regard realism about the wave function\index{wave function} as the most promising stance. Since we do not assume familiarity with phenomenology\index{phenomenology} on the part of our readers, \hyperref[sec3]{Section 3} offers a concise introduction to some of the central tenets of phenomenological philosophy. Building on this, \hyperref[sec4]{Section 4} takes a closer look at the London–Bauer Interpretation (hereafter \textit{LBI\index{London-Bauer-Interpretation}})—the most elaborate attempt to interpret quantum theory informed by the then-burgeoning phenomenological movement. In \hyperref[sec5]{Section 5}, we consider what adopting LBI\index{London-Bauer-Interpretation} would mean for WFR. 

\section{Realism about the wave function}\label{sec2}

\subsection{The wave function and the measurement problem}

The wave function\index{wave function} occupies a central place in quantum mechanics: it is taken to represent the state of a quantum system and to evolve in time according to the Schrödinger equation---or so it is often said. Yet this seemingly straightforward picture immediately encounters a well-known tension. As \cite{Maudlin95} famously noted, the following three theses are mutually inconsistent: 
\begin{enumerate}[(i)]
\item The wave function\index{wave function} fully describes the state of the system.
\item The wave function\index{wave function} always evolves according to the Schrödinger equation.
\item We typically observe definite outcomes, not superpositions.
\end{enumerate}
At least one of these claims must give way. This is the so-called \textit{measurement problem\index{measurement problem}}. It follows directly from the linearity of the Schrödinger equation, which entails that the wave function\index{wave function} of a system will in general evolve into a superposition\index{superposition} of distinct states---something that appears incompatible with the definitness of our measurements and everyday experience. Different interpretations of quantum mechanics resolve the tension by denying different members of this triad. Hidden-variable theories such as Bohmian mechanics (BM) reject (i); textbook quantum mechanics and objective-collapse theories, such as GRW, reject (ii); the many-worlds interpretation (MWI) rejects (iii).

BM rejects (i) by introducing additional variables, namely point particles with definite positions. The complete state of a system is therefore given not only by the wave function\index{wave function} but also by the actual configuration of these particles. The wave function\index{wave function} itself continues to evolve according to the Schrödinger equation, yet BM adds a second deterministic law---the guiding equation---that specifies how the particle configurations evolve over time. Taken together, these equations define a dynamics that departs markedly from textbook quantum mechanics and alters its formalism significantly. 

MWI, by contrast, introduces no new variables or equations. It endorses (i) and (ii) but rejects (iii), maintaining that whenever a quantum event occurs, reality branches so that every possible outcome is realized in a separate world. In terms of the formalism, MWI is the most economical interpretation of quantum mechanics: it leaves the mathematical structure entirely intact. In terms of ontology, however, it is far from economical since the theory entails the existence of infinitely many parallel worlds---or, more precisely, infinitely many \textit{in principle unobservable} entities---a feature that many find deeply troubling.

Textbook quantum mechanics and GRW both reject (ii), but in very different ways. GRW \textit{modifies} the Schrödinger equation by introducing spontaneous collapses\index{collapse} of the wave function\index{wave function}, which render its dynamics stochastic and non-linear. Since both GRW and Bohmian mechanics modify the formalism of quantum mechanics, it is often said that, strictly speaking, they are not \textit{interpretations} of quantum mechanics but \textit{rival theories}. This is not without cost. It seems counter-intuitive to modify a theory that is widely regarded as one of the greatest achievements of modern science. To add insult to injury, as \cite{Wallace2023} notes, there currently exists no version of Bohmian mechanics or GRW that is consistent with special relativity. In this sense they are not so much rival theories of standard quantum mechanics as \textit{empirically inferior alternatives}. 

Textbook quantum mechanics, on the other hand, rejects (ii) through the infamous projection or collapse postulate. According to this postulate, the wave function\index{wave function} evolves according to the Schrödinger equation except when a measurement takes place. Upon measurement, the wave function\index{wave function} is assumed to collapse\index{collapse} into one of its eigenstates, and a definite outcome is observed. 

The obvious follow-up question, then, is why the wave function\index{wave function} collapses. What, if anything, is so special about measurement? This question is particularly pressing if one interprets the wave function\index{wave function} as an ontologically real entity and the collapse\index{collapse} as a physical process. For if there truly exists a physical object called the wave function\index{wave function} that evolves continuously according to our quantum equation of motion---that is, the Schrödinger equation---then it remains mysterious why a second, discontinuous type of dynamics should suddenly take over whenever a measurement occurs. 

For this reason, it is often said that the main proponents of the so-called Copenhagen interpretation, especially Bohr\index{Bohr, Niels} and Heisenberg\index{Heisenberg, Werner}, did not regard the collapse\index{collapse} of the wave function\index{wave function} as a physical process at all, but rather as a change in our \textit{knowledge} or \textit{information}. This is how the view is frequently presented in the philosophy of physics literature, although the historical reality is considerably more complex.\footnote{It would lead us too far afield to enter a detailed discussion of the history of quantum physcis. In the context of this chapter, it suffices to note that, as recent scholarship has shown, the idea of a unified Copenhagen Interpretation that sailed under the same banner ever since Bohr\index{Bohr, Niels} first introduced the concept of complementarity in 1927 is a mythological construction. Rather, what became known as the Copenhagen Interpretation was an assemblage of more or less diverging ideas from Bohr\index{Bohr, Niels}, Heisenberg\index{Heisenberg, Werner}, Dirac, von Neumann, and others \citep[]{Howard2004}.} In any case, if one takes a further step and interprets the wave function\index{wave function} not as representing an external system but as encoding a subject's belief or information, the measurement problem\index{measurement problem} dissolves entirely. Even some strong opponents of subjective interpretations acknowledge this much.

\begin{quote}
Any approach according to which the wave function\index{wave function} is not something real, but represents a subjective information, explains the collapse\index{collapse} at quantum measurement perfectly: it is just a process of updating the information the observer has \cite[17]{Vaidman2014}.
\end{quote}

It is noteworthy that many physicists are relatively comfortable with the idea that science cannot deliver a purely objective description of reality and that it necessarily involves subjective or operational notions such as \enquote{information,} \enquote{experience,} or \enquote{measurement.} Philosophers of physics, by contrast, often regard the very presence of these concepts in the foundations of quantum mechanics as problematic. As Maudlin puts it, \enquote{A precisely defined physical theory […] would never use terms like \enquote{observation,} \enquote{measurement,} \enquote{system,} or \enquote{apparatus} in its fundamental postulates. It would instead say precisely what exists and how it behaves} \cite[5]{Maudlin2019}.

The so-called \enquote{measurement problem\index{measurement problem},} then, is not a matter of potential internal inconsistencies within textbook quantum mechanics---though such tensions certainly exist. The deeper difficulty lies in the very appearance of the concept of measurement in the axioms of our theory. Many philosophers of physics regard this as unsatisfactory: a fundamental physical theory, they argue, should not have to invoke the act of measurement at its core. The appeal of interpretations such as BM, MWI, and GRW is precisely that they promise to eliminate the notion of measurement at the fundamental level. The cost, however, is high: one must either modify the formalism or accept the existence of many worlds. 

A widespread assumption in the foundations of physics is that adopting a realist interpretation of quantum mechanics naturally commits one to taking the wave function\index{wave function} as physically real. The analogy with classical mechanics is often invoked at this juncture: just as the positions and momenta of point particles not only describe the state of the system but represent physically real quantities that evolve according to the equations of motion (for instance, Hamilton's equations), so too the wave function\index{wave function} in quantum mechanics is taken to represent a physically real entity evolving in time according to the Schrödinger equation.

That matters are not so straightforward becomes clear when we recall that in BM the state of a system is not described by the wave function\index{wave function} alone, but by the wave function\index{wave function} together with the configuration of the point particles. In this context, it can be more natural to interpret the wave function\index{wave function} \textit{nomologically} rather than \textit{ontologically}---that is, as a quantity that guides the particles not by exerting physical influence but in a law-like manner \citep{GoldsteinZanghi}. In other words, BM does not straightforwardly lead to WFR. 

Another worry arises when we recall that even in classical mechanics we cannot read an ontology directly off the formalism, at least not as easily as it is often suggested \citep{North2022,Wilson2013}. Consider a classical system of $N$ particles. It is customary to say that the \enquote{state of the system is represented by $N$ points $X_1,...X_N$, in three-dimensional Euclidean space} \cite[68]{Wallace2021}. Yet this is only one possible mathematical representation. As Wallace notes, \enquote{there is another way to represent this theory. We can define the configuration space as the product of $N$ copies of Euclidean three-space. Each N-tuple of points $(X_1,...X_N)$ now corresponds to a single point in this 3$N$-dimensional space} \cite[68]{Wallace2021}. Wallace offers this example specifically as an argument against WFR---a point to which we will return below.

\subsection{Reifying the wave function}

The discussions so far already point to a deeper tendency in many realist approaches to quantum mechanics: the inclination to treat elements of the formalism as if they directly corresponded to physical entities. This move---common in the history of science and particularly tempting in the case of the wave function\index{wave function}---invites a crucial distinction.

The first thing to note is the difference between two claims: that the wave function\index{wave function} represents or describes the state of a physically real system, and that the wave function\index{wave function} \textit{itself} is physically real. The latter view holds that the mathematical object we call the wave function\index{wave function} represents a physically existing entity \enquote{out there} that evolves according to a dynamical law such as the Schrödinger equation. On this view, this entity is not a particle, nor an ordinary field defined over three-dimensional space, but something of a new kind---an entity properly called a wave function\index{wave function}. To maintain that the mathematical formalism of quantum theory refers to such an independently real wave function\index{wave function} is to \textit{reify}, or \textit{objectify}, the wave function\index{wave function} itself.

In straightforward terms, this position would be called \textit{wave function realism}. However, the label is already reserved for a specific way of reifying the wave function\index{wave function}: the view that the wave function\index{wave function} is a physically real field defined on configuration space and that this configuration space is itself ontologically fundamental.\footnote{Wave function realists sometimes say that the fundamental space in which the wave function\index{wave function} lives is a space that \enquote{has the structure of a configuration space.} Why not simply say that it \textit{is} configuration space? The rationale for this phrasing is to emphasize that, for the wave function realist, the configurations of particles are not fundamental (see \citealt[41f.]{Ney2021}).} As should be clear from the preceding discussion, and as several authors have pointed out \citep{Myrvold2015,Wallace2021,Chen2019}, this terminology can be misleading. One can, for example, hold that both the wave function\index{wave function} and configuration space are real but not fundamental \cite{Myrvold2015}. Or one might propose that the wave function\index{wave function} is real, yet should not be understood as a field on configuration space at all, but as a physically real vector in a physically real Hilbert space---a position its advocates call \enquote{Mad-Dog Everettianism} \citep{CarrollSingh}. Another possibility is to regard the wave function\index{wave function} not as a field on multi-dimensional configuration space but a \textit{multi-field} defined on ordinary three-dimensional space \citep{HubertRomano}. Hence in a more careful terminology, we can distinguish at least three ways of reifying the wave function\index{wave function}: \textit{configuration space realism}\index{configuration space realism}, \textit{Hilbert space realism}, and \textit{multi-field realism}. 

Among these options, configuration space realism\index{configuration space realism} remains the most widely held and extensively discussed. In the next subsection, we will introduce this position in proper mathematical form and then discuss its main motivations and challenges. For ease of presentation, we will follow standard usage and treat the terms “wave function realism" and “configuration space realism" synonymously.

\subsection{WFR: Motivations and objections}

For an $N$-particle system, the \emph{configuration} of the system is given by the complete specification of all particle positions at a given instant. The collection of all such possible configurations defines the system's \emph{configuration space}. For $N$ particles in ordinary three-dimensional physical space, a single point
$$
(q_1, q_2, \dots, q_{3N}) \in \mathbb{R}^{3N}
$$
specifies a complete configuration of the system. The corresponding configuration space therefore has dimensionality $3N$.

In quantum mechanics, the wave function\index{wave function} is introduced as a complex-valued function defined on configuration space,
$$
\Psi: \mathbb{R}^{3N} \longrightarrow \mathbb{C},
$$
which assigns to each configuration a complex number $\Psi(q_1, q_2, \dots, q_{3N})$. The corresponding probability density for finding the system in that configuration is given by
$$
P(q_1, q_2, \dots, q_{3N}) = |\Psi(q_1, q_2, \dots, q_{3N})|^2,
$$
where the wave function\index{wave function} satisfies the usual normalization condition,
$$
\int_{\mathbb{R}^{3N}} |\Psi(q_1, q_2, \dots, q_{3N})|^2 \, d^{3N}q = 1.
$$

Given that there are estimated to be around $10^{80}$ particles in the universe, WFR implies that the fundamental space we inhabit---namely, the configuration space in which the universal wave function\index{wave function} is defined---has a dimensionality of roughly $3 * 10^{80}$. This immediately raises the question: what, then, becomes of the \enquote{ordinary} three-dimensional space of our everyday experience? Proponents of WFR have offered two answers. Either declare that the familiar three-dimensional world is “flatly illusory" \cite[277]{Albert1996} or regard it as an emergent phenomenon \citep{Albert2015,Albert2019}. 

One might be tempted to respond that if an interpretation of quantum mechanics entails that our everyday experience is illusory in any substantial sense, so much worse for that interpretation. As Bradley Monton puts it, “our everyday commonsense constant experience is such that we’re living in three spatial dimensions, and nothing from our experience provides powerful enough reason to give up that prima facie obvious epistemic starting point” \cite[154]{Monton2013}. Peter Lewis, similarly, argues “for the converse of Albert’s initial position; the world really is three-dimensional, and the 3N-dimensional appearance of quantum phenomena is the theoretical analog of an illusion” \cite[124]{Lewis2013}. 

From a phenomenological point of view, this issue strikes at the very heart of the epistemic foundations of science. If one takes seriously the idea that experience is the primary source of justification, then it seems blatantly absurd to interpret empirical science in a way that undermines the cognitive legitimacy of experience itself---since doing so would ultimately threaten to undermine the sciences as such.\footnote{Following Jeffrey Barrett \cite[116]{Barrett1999}, Huggett and Wüthrich define \enquote{a theory to be empirically incoherent in case the truth of the theory undermines our empirical justification for believing it to be true} \cite[277]{HuggettWuethrich2013}. Hence, if our everyday experiences are the epistemic foundation of the physical sciences, but physics is interpreted as revealing that experiences are fundamentally illusory, this interpretation of physics is in danger of being empirically incoherent.} More on this in Section~\ref{sec3}.

Alternatively, one may argue that our three-dimensional space is indeed real but \textit{emergent}. Yet this move raises its own challenge: one must explain how a space with the structure of ordinary three-dimensional Euclidean space can emerge from a space with the vastly different structure of configuration space. The difficulty is compounded by the fact that it remains unclear what kind of structure proponents of WFR ascribe to the space they take to be fundamental \citep{Myrvold2025}.

Let us now consider why one might be drawn to WFR in the first place. The most immediate motivation, already mentioned above, is that WFR appears to follow naturally from adopting a realist attitude toward the quantum formalism itself. As Wallace notes, when we present the wave function\index{wave function} for an $N$-particle system in the standard way, \enquote{[w]ave function realism […] then seems to follow straightforwardly from the formalism} \citep{Wallace2021}. Likewise, as \cite{Myrvold2025} observes, proponents of WFR often suggest that their position is the most direct consequence of realism about quantum mechanics more generally (see also \citealt[714]{Lewis2004}).\footnote{This applies, for instance, to the work of Albert. Ney, by contrast, is more cautious, arguing that while Bohmian mechanics is consistent with WFR, it is not a particularly natural fit \cite[24, 43]{Ney2021}. However, she also maintains that WFR offers the natural interpretitive framework for proponents of MWI and GRW.} However, there are at least four problems with this line of reasoning.\footnote{A further, perhaps even more serious, problem for the claim that the wave function\index{wave function} and its configuration space are fundamental arises from the fact that non-relativistic quantum mechanics is \textit{not} itself a fundamental theory. Matters become considerably less straightforward once we move to relativistic quantum mechanics or quantum field theory. For details, see \citep{Myrvold2015,Wallace2021}. In what follows, we therefore focus on the interpretitive question of whether WFR remains plausible \textit{if} non-relativistic quantum mechanics were taken to be fundamental.}

First, why should we be realist and fundamentalist about configuration space rather than, for instance, about Hilbert space? Every possible pure state of a quantum system corresponds to a vector in complex Hilbert space. On a mathematically fundamental level, then, wave functions\index{wave function} \textit{are} vectors in Hilbert space---hence the familiar claim that \enquote{[w]ave functions live in Hilbert space} \cite[94]{Griffiths2018}. Why, then, should configuration space rather than Hilbert space be regarded as the physically real arena?

For many, the answer seems to be that Hilbert space is simply too abstract to qualify as a serious candidate for physical reality. As Wallace notes, “[v]ery few people are willing to defend Hilbert space realism in print” \cite[216]{Wallace2013}. One exception is \cite{CarrollSingh}, who call their view \enquote{Mad-dog Everettianism}. Yet, if the perceived abstractness of Hilbert space counts as a reason to reject Hilbert space realism, then---by the same measure---this should also count against WFR. To the extent that our interpretation of quantum mechanics can avoid reifying such abstract mathematical spaces, so much the better.\footnote{Perhaps the first objection to something like WFR can be found in Weyl’s\index{Weyl, Hermann} 1934 text \textit{Mind and Nature}: “One could devise the following expedient and say: The wave field governed by strict laws is the real thing. How amiss that is is shown right away by the following consideration: If we are dealing with two electrons, we can inquire as to the probability that the one is found at the place with the coordinates ($x_1, y_1, z_1$) the other at the place with the coordinates ($x_2, y_2, z_2$). The wave function\index{wave function} $\psi$ determining this probability must therefore be a function of the two positions in space $x_1, y_1, z_1; x_2, y_2, z_2$, or the wave in this case does not extend in the usual three-dimensional space but in a six-dimensional space. The more particles are added, the higher the number of dimensions of the space rises in which the de Broglie wave process takes place. This alone should show sufficiently that the wave field is only a theoretical substructure" \cite[147f.]{Weyl2009}.}

Second, any vector in Hilbert space can be expressed in multiple representations or bases. Instead of working in the position basis, one might equally choose the momentum basis, in which the wave function\index{wave function} is defined over momentum space. Mathematically, this switch of representation is entirely straightforward---the two are related by a Fourier transform. Yet conceptually, the situation is far less innocent: if the proponent of WFR identifies the wave function\index{wave function} with a physically real entity, must she then regard \textit{all possible} representation spaces as real? The formalism alone offers no guidance on which basis, if any, corresponds to the \enquote{true} underlying space. This ambiguity complicates any straightforward realist reading of the wave function\index{wave function}.

Third---and perhaps more pressing---every pure state of a quantum system corresponds to a Hilbert space, but this correspondence is \textit{not} unique \cite[note 7]{Ismael2025}. Any wave function\index{wave function} multiplied by a complex number of absolute value 1 represents precisely the same state. As \cite{Myrvold2025} points out, however, proponents of WFR typically speak as though there was such a thing as \textit{the} wave function\index{wave function}. We are invited, for instance, to be realists about \textit{the} universal wave function\index{wave function} that describes the state of the entire universe. Yet which function is that? If the realist's talk of \textit{the} wave function\index{wave function} is taken literally, this marks a radical departure from standard quantum theory, according to which no unique function corresponds to a given state. As Myrvold emphasizes, \enquote{there is no such thing as \textit{the} wave function\index{wave function} that represents a quantum state} \citep{Myrvold2025}. From an intuitive standpoint, it seems preferable for an interpretation to preserve, rather than revise, a formalism that has proved to be remarkably successful.\footnote{One could argue that all the wave function realist has to do to avoid such problems is to replace wave functions with \textit{rays}. A ray is an equivalence class of state vectors differing only by a global phase $e^{i\theta}$. While different wave functions in the same ray represent the same physical state, the ray represents that state uniquely. Thus, it may seem attractive for the realist to move their ontological commitment to rays. However, this move comes at a significant cost: it abandons the core tenet of configuration space realism. Rays are not functions on configuration space but abstract elements of Hilbert space. Consequently, the resulting Hilbert space realism abandons both (i) the idea that the fundamental entities are field-like quantities and (ii) the idea that these entities evolve on configuration space. In other words, the move to rays would not constitute a recognizable form of wave function realism. See \cite[Section 6]{Ney2023} for why she prefers WFR to Hilbert space realism.}

Fourth, even apart from the internal difficulties just mentioned, there are general philosophical reasons for caution when it comes to reifying the mathematics of our physical theories---and these reasons acquire particular force if one has sympathies with the phenomenological tradition. To put a long story short: From a phenomenological standpoint, mathematical formalism is a method of representation and sometimes even a cognitive lens for constitution\index{constitution}; but it is not a window onto being itself. As Husserl\index{Husserl, Edmund} warned, the danger is to \enquote{take for true being what is actually a method} \cite[51]{Husserl1970}. We will return to the broader implications of this critique of reification of mathematics in Section~\ref{sec3}. 

For now, it suffices to emphasize that what makes the project of reifying the mathematics particularly problematic is that in our best physical theories we never have a unique mathematical representation of reality.\footnote{“[E]very theoretical physicist who is any good knows six or seven different theoretical representations for exactly the same physics. He knows that they are all equivalent, and that nobody is ever going to be able to decide which one is right at that level, but he keeps them in his head, hoping that they will give him different ideas for guessing” \cite[168]{Feynman1965}.} This non-uniqueness has become a central theme, for instance, in the philosophy of gauge theories (\citealt[Section 15.3.1]{Berghofer2022} and \citealt{Berghofer-et-al2023}) and dual theories (\citealt{DeHaroButterfield}). In the present context, WFR invites comparison with a parallel move in classical mechanics that, as Wallace emphasizes, would seem plainly absurd. Instead of regarding individual particles as the fundamental objects moving in three-dimensional Euclicean space, one could \textit{formally} recast classical $N$-particle mechanics as describing a single point moving in a 3$N$-dimensional configuration space. Yet no one takes this reformulation to reveal the true physical arena of classical physics. Wallace concludes:

\begin{quote}
When physical theories are presented to us as formulated on spaces with much more structure than Euclidean space, we should not rush to interpret them as \textit{physical} spaces, rather than as mathematical devices to encode information about the physical state. \cite[69]{Wallace2021}
\end{quote}

We agree. Yet it is important to note that Alyssa Ney---alongside Albert, the most influential proponent of WFR---explicitly denies that a straightforward reading of the quantum formalism is what motivates her position. Drawing on Maddy's classical discussion of naturalism in science \citep{Maddy1992}, Ney emphasizes that it \enquote{is a common view in post-Quinean metaphysics of science} that we must not \enquote{read the ontology of a theory off of a mathematical description of that theory} \citep{Ney2023}.\footnote{For systematic similarities between Maddy's and Husserl\index{Husserl, Edmund}'s approach to science, see (\citealt{Hartimo2021,Hartimo2020}).} 

We share this caution, though it seems that this insight is still far from firmly established in contemporary philosophy of physics. Moreover, we note that Ney dismisses this “prima facie case" for WFR only in the sense that it does not allow us to prefer WFR over other \textit{realist} interpretations of the wave function\index{wave function} such as, e.g., the multi-field view. Her own question, ultimately, is different: \textit{granting} a realist interpretation of the wave function\index{wave function}, why should we prefer WFR over competing realist accounts such as the multi-field view? Her answer is what she calls the \enquote{argument from separability and locality}.

Given our everyday experience, it is natural to assume that a physical object is affected only by its immediate surroundings. We do not, after all, encounter \enquote{spooky action at a distance.} If Ney is right that WFR distinguishes itself from its realist rivals in being the only view to offer a local description of reality, this would seem to count in its favour. Yet WFR achieves locality only with respect to configuration space. It remains non-local where it seems to matter most, namely in the ordinary three-dimensional space of experience. Since our intuitions about causation and locality are shaped by what appear to us as objects and events in three-dimensional space, it is doubtful whether preserving locality in an abstract, high-dimensional configuration space can carry the interpretitive weight Ney assigns to it. If anything, it would seem an advantage for an intepretation to retain locality in the space of lived and observed phenomena. But only a non-realist intepretation of the wave function\index{wave function} appears capable of doing so---or so it seems.

The perhaps most consistent non-realist interpretation compatible with locality in our lifeworld is QBism\index{QBism}\index{QBism\index{QBism}}, according to which the wave function\index{wave function} represents the subjective degrees of belief of an experiencing agent. On this view, quantum mechanics is not a description of an objective reality but a tool for forming consistent expectations about future experience. In this respect, QBism\index{QBism} already points toward a \textit{first phenomenological option}: it relocates the centre of gravity of the theory from an objective world of things to the domain of lived experience. Its focus on the first-person perspective and its refusal to reify the mathematical formalism resonate strongly with the phenomenological insight that scientific representation is always and necessarily mediated by the structures of subjectivity (\citealt{Bitbol2020,Tremblaye2020,BerghoferWiltsche2023}).

According to how QBism\index{QBism} is often portrayed in the literature, it resolves the problem of non-locality in a single stroke—but, according to some critics, at a steep price. From the critics’ point of view, QBism\index{QBism} appears to embrace a form of radical subjectivism: the wave function\index{wave function} no longer represents \textit{reality} in any straightforward or substantial sense, but instead encodes the agent’s personal degrees of belief. In this light, statements by QBists themselves---such as Fuchs and  ollaborators’ claim that \enquote{quantum mechanics should be viewed as a decision theory} \citep[8]{debrota2024} rather than a theory about the world---do little to dispel the impression that something crucial is lost: the sense that quantum mechanics, however indirectly or non-objectivistically, still is a theory about reality or at least tells us something about the structure of reality itself.
\footnote{Importantly, QBists themselves explicitly reject the labels of subjectivism and anti-realism and emphasize that although they are anti-realist about the wave-function, they still believe that they are part of a realist research project in a broader sense. Specifically, they believe that there are ontological lessons to be learned from a QBist interpretation of quantum mechanics. Perhaps the most important lesson being “that reality is more than any third-person perspective can capture” \cite[113]{Fuchs2017}.}
We will return to this issue in Section~\ref{sec5}.

This brings us to a second phenomenological option. If QBism\index{QBism}, at least as it is often portrayed, risks collapsing into an overly subject-centered picture, it is worth recalling that long before the debates about QBism\index{QBism} began, the phenomenological tradition had already proposed a framework that acknowledges the perspectival rooting of science in the observer while insisting that this very rooting is what makes the constitution\index{constitution} of objectivity possible in the first place. 

The appeal of such an alternative becomes clearer once we recall what motivates many contemporary realists. For authors like Ney, quantum mechanics is \enquote{intended to
be descriptive of an object or objects that exist \textit{independently of us or any other observer}} (\citealt[530]{Ney2012}; our emphasis), that is, it \enquote{describes what the world is like, \textit{independent of us as observers}} (\citealt[3110]{Ney2015}; our emphasis). The task of interpretation is therefore to understand quantum mechanics in a way that brings the theory into line with this aim and to reconstruct reality purified of any subjective contribution. The phenomenological tradition, by contrast, begins from the opposite insight: that the very distinction between world and observer is itself an achievement—something constituted through the structures of experience. Before we can appreciate how this insight reshapes the interpretation of quantum mechanics—and before turning to the London–Bauer interpretation, which develops this idea most explicitly—we must pause and say something about phenomenology\index{phenomenology} itself.

\section{phenomenology\index{phenomenology} for beginners}\label{sec3}

\subsection{From the structures of experience to the epistemic role of the life-world\index{life-world}}

phenomenology\index{phenomenology}---understood in the Husserlian tradition inaugurated at the start of the twentieth century---is the study of the structure of consciousness, and of conscious experience in particular. More precisely, it is a \textit{descriptive} and \textit{eidetic} inquiry. It is descriptive in that it proceeds from a first-person perspective: the aim is not to identify, say, which neural circuits are activated during a visual episode, but to clarify what it is like for a subject to undergo that kind of experience. It is eidetic in that it seeks the invariant structures that hold for any possible experiencing subject, not idiosyncrasies of you or me. A paradigmatic structural invariant is \textit{intentionality}: every conscious experience is of or about something. phenomenology\index{phenomenology} thus investigates the \textit{intentional structure} of experience---how experiences are directed toward their objects and in virtue of what this directedness obtains.

A crucial phenomenal feature that distinguishes, say, perceptual experiences from mere thoughts, beliefs, or hopes is that the former exhibit what Husserl\index{Husserl, Edmund} calls \enquote{originary givenness.} In contemporary epistemology, this feature is often discussed under the heading of \enquote{presentational phenomenology\index{phenomenology}} (\citealt{Chudnoff2013}). Here “phenomenology\index{phenomenology}” refers to the \textit{phenomenal character} of experience, that is, the character of \enquote{what it is like subjectively to undergo the experience} \cite[Section 1]{Tye2015}. Perceptual experience possesses the character of originary givenness insofar as it does not merely represent its objects but \textit{presents} them \enquote{in the flesh} \cite[458]{Husserl1984}, as \enquote{bodily present} \cite[14]{Husserl1973b}. In other words, there is a clear \textit{phenomenal contrast} between thinking about my friend, believing or hoping that I will see her tomorrow, and actually seeing her. 

Importantly---and this marks a decisive difference between Husserl\index{Husserl, Edmund}, who has been described as a radical empiricist \citep[]{Wiltsche2025}, and standard empiricists---the character of originary givenness is not confined to perceptual experience. Other types of experience exhibit it as well, including \textit{a priori} intuitions such as logical and mathematical insights and even certain forms of introspection. There is, again, a clear phenomenal contrast between merely recalling that two is the only even prime number, or believing this on the basis of memory, and actually \textit{intuiting} the proposition as evident. While it is true that mathematical intuitions do not present their objects \enquote{in the flesh,} they nonetheless exhibit a genuinely presentational phenomenology---one that contrasts distinctively with the mere holding of a belief.\footnote{\enquote{[P]erception [...] makes an individual object given originarily in the consciousness of seizing upon this object 'originarily,' in its 'personal' selfhood. In quite the same manner intuition of an essence is consciousness of something, an 'object,' a Something to which the intuitional regard is directed and which is 'itself given' in the intuition} \cite[9f.]{Husserl1982}. In this terminology, “intuition of an essence” corresponds to what we call an a priori intuition. In contemporary epistemology, several positions share with Husserl\index{Husserl, Edmund} the idea that a priori intuitions have a presentational phenomenology\index{phenomenology} analogously to perceptual experiences (see, e.g., \citealt{Bengson2015,Berghofer2022,Chudnoff2013,Church2013}).}

A further feature brought to light by the phenomenological analysis of intentional experience is that perceptual acts not only exhibit originary givenness but also a \textit{horizon of co-givenness}. Experiences are, in this sense, horizonally structured: they always extend beyond what is directly presented. Experiencing is, as Husserl\index{Husserl, Edmund} puts it, always an \enquote{\textit{intending-beyond-itself}} \cite[46]{Husserl1960}. Consider the simple example of looking at a coffee cup. At first glance, what appears to you is a three-dimensional object in space. Yet a closer examination reveals that what is strictly sensuously given is not the cup as a whole, but only \textit{one profile}---its currently visible front side. Of course, you could wander around and make the current back side the new front side, and vice versa. But this doesn’t change the fact that the cup is always given in perspectives and that, more generally, the objects of perceptual experiences always and necessarily have more parts, functions, and properties than can be actualized in one single intentional act. As Madary puts it, \enquote{Visual perception is an ongoing process of anticipation and fulfillment} \cite[3]{Madary2017}. When you look at the cup, you already anticipate how it will feel to the touch, how it will appear from another angle, or how it will sound when set down—all anticipations that may be fulfilled in new acts of perception, which in turn generate further anticipations.\footnote{In other words, our current perceptions are embedded into a manifold of possible “perceptions that we could have, if we actively directed the course of perception otherwise: if, for example, we turned our eyes that way instead of this, or if we were to step forward or to one side, and so forth” \cite[44]{Husserl1960}. Thus, perception is a composite of interwoven \enquote{fulfilled and unfulfilled intentions} \cite[221]{Husserl2001b}.}

In other words, a close look at how physical objects appear to us reveals that our experiences always \enquote{transcend} or \enquote{go beyond} what is directly sensuously given to us. There is a clear distinction between what is \textit{intended} in a perceptual act—for instance, that there is a coffee cup before you—and what is \textit{sensuously given}, namely the object’s facing side with its momentarily visible features. For the phenomenologist, this discrepancy is \textit{not} a flaw to be explained away. Instead, the fact that our perceptual intentions always transcend the sphere of direct givenness is to be treated as a phenomenologically discoverable feature of experience itself. The perspectival and horizonal structure of experience are therefore not imperfections of human perception but constitutive features of how the world is disclosed to us.\footnote{\enquote{Necessarily there always remains a horizon\index{horizon} of determinable indeterminateness, no matter how far we go in our experience, no matter how extensive the continua of actual perceptions of the same thing may be through which we have passed} \cite[95]{Husserl1982}.}

For Husserl\index{Husserl, Edmund}, such phenomenological investigations have decisive \textit{epistemological} significance. As he succinctly puts it, “No epistemology without phenomenology\index{phenomenology}” \cite[217]{Husserl1984b}. Certain forms of experience—such as perception, introspection, or mathematical intuition—are not only phenomenally distinctive in being \textit{originary presentive} (as opposed to mere beliefs or hopes), but also serve as sources of \textit{epistemic justification}.\footnote{For more details on Husserl\index{Husserl, Edmund}’s (moderate) foundationalism and his approach to the justificatory force of experience, see \cite[Chapters 5\&7]{Berghofer2022}. For Husserl\index{Husserl, Edmund}, the question of how subjective experiences can be a source of objective knowledge is the perhaps most fundamental epistemological question (see Melle in \cite[p.xxxi]{Husserl1984b}). His phenomenological-internalist approach that certain experiences justify by virtue of their distinctive phenomenal character is shared by several contemporary epistemologist. For an overview, see \cite{Berghofer2020}. For a defense of phenomenological internalism, see \cite[Chapter 4]{Berghofer2022}.} Suppose you merely believe or hope that the window in your room is closed, and then go and see that it is: the act of seeing is not only phenomenally distinct from belief or hope, it also \textit{confers justification}. Beliefs and hopes do not justify non-inferentially; perceptual experiences do. For Husserl\index{Husserl, Edmund}, this is no accidental overlap between phenomenology\index{phenomenology} and epistemology. Rather, every originary presentive experience justifies precisely \textit{in virtue of} its distinctive presentational phenomenology\index{phenomenology} \cite[36f., 44]{Husserl1982}. In this sense, all knowledge and epistemic justification can ultimately be traced back to originally presentive experience.

From this vantage point, the problem of modern science appears in a new light. If all theoretical knowledge must, in the last instance, be grounded in originary modes of givenness, then the increasing abstraction of the sciences raises a pressing question: how can highly mathematized theories—whose objects are never directly given—remain anchored in the sphere of experience from which their sense and justification ultimately derive? It is precisely this question that motivates Husserl\index{Husserl, Edmund}’s analysis in the \textit{Crisis} \citep[]{Husserl1970}. The issue is not with mathematization\index{mathematization} \textit{per se}—which Husserl\index{Husserl, Edmund} recognizes as an extraordinary cognitive achievement—but with a gradual forgetting of its experiential origin. When the formal constructions of science are taken to depict reality \textit{itself}, the relation between theoretical representation and the world of lived experience is inverted. The very practices that were meant to clarify experience come to obscure it.

To be sure, Husserl\index{Husserl, Edmund} deeply admired the accomplishments of Galileo and his successors: that we can describe the world so effectively in mathematical terms is, for him, an astonishing accomplishment that calls for serious philosophical reflection. What he objects to is not mathematization\index{mathematization} itself, but the way its success has been \textit{interpreted}. More specifically, the problem arises when mathematics is reified—when the formal models designed to represent reality are mistaken for reality itself, and the world of everyday experience is dismissed as merely illusory. As Husserl\index{Husserl, Edmund} writes:

\begin{quote}
But now we must note something of the highest importance that occurred even as early as Galileo: the surreptitious substitution of the mathematically substructed world of idealities for the only real world, the one that is actually given through perception, that is ever experienced and experienceable—our everyday life-world\index{life-world}. This substitution was promptly passed on to his successors, the physicists of all the succeeding centuries. \cite[48f.]{Husserl1970} 
\end{quote}

Husserl\index{Husserl, Edmund} describes mathematical models as a \enquote{garb of ideas} through which we clothe reality. Yet, he warns, this garb is often mistaken for reality itself: \enquote{It is through the garb of ideas that we take for \textit{true being} what is actually a \textit{method}---a method which is designed for the purpose of progressively improving, \textit{in infinitum}, through \enquote{scientific} predictions, those rough predictions which are the only ones originally possible within the sphere of what is actually experienced and experienceable in the lifeworld} \cite[51f.]{Husserl1970}.

To summarize, experience constitutes the most basic source of epistemic justification. Everything we know---both about the world and through science---can, in the end, be traced back to epistemically foundational experiences. The world of everyday experience is therefore more fundamental, in epistemic terms, than the \enquote{world of science.} For Husserl\index{Husserl, Edmund}, the life-world\index{life-world} is the \enquote{forgotten meaning-fundament of natural science} \cite[48]{Husserl1970}.  

In the context of wave function realism, this perspective entails a characteristic phenomenological skepticism toward reifying the wave function\index{wave function}. To treat the wave function\index{wave function} as the truly real entity is to mistake a method of predicting reality for reality itself—and, worse still, to adopt an interpretation that is self-undermining insofar as it renders the very foundation of scientific knowledge---the life-world\index{life-world} of experience---illusory.

\subsection{On the subject-object relationship and subjectivity in science}

At this stage, it is useful to clarify what is---and what is not---at stake in Husserl\index{Husserl, Edmund}’s
critique of mathematization\index{mathematization}. We can distinguish two levels of argument. The first is a
modest \textit{interpretational} claim: mathematical objects in physical theories
should not be reified. As discussed above, Husserl's\index{Husserl, Edmund} warning against \enquote{taking for true
being what is actually a method} (\citealt[51]{Husserl1970}) can be read as a call for
epistemic restraint, much in the spirit of van Fraassen’s constructive empiricism \citep[]{Wiltsche2012}. On this view, scientific models are idealized \textit{garbs of ideas}—conceptual tools whose justified scope extends only as far as their empirical fit with the world of experience. To project these models beyond that sphere, by treating their abstract structures as literal constituents of reality, is to mistake a method of representation for an ontology.

The second, stronger, claim is \textit{methodological}. It asserts not only that we should resist reifying mathematical entities, but that the very practice of science is inseparable from the perspective of the subject who experiences and theorizes. The natural sciences, on this view, can never achieve a purely objective, third-person standpoint. As Merleau-Ponty\index{Merleau-Ponty, Maurice} famously put it, “Everything that I know about the world, even through science, I know from a perspective that is my own or from an experience of the world without which scientific symbols would be meaningless. The entire universe of science is constructed upon the lived world”
\cite[lxxii]{MerleauPonty2012}. Several contemporary phenomenologists share this conviction (see, e.g., \citealt[54]{Zahavi2019}).

The step from the interpretational to the methodological claim marks an important shift: it transforms Husserl\index{Husserl, Edmund}’s epistemic caution into a positive thesis about the conditions of possibility of scientific knowledge. Far from being a limitation, subjectivity becomes the very medium through which objectivity is constituted. This marks a decisive departure from the idea of science as a view from nowhere: the phenomenological insight is that any account of the world, including the scientific one, must begin from within the horizon\index{horizon} of lived experience itself. In this sense, Husserl’s critique of mathematization\index{mathematization} is not only a call for epistemic modesty but also an invitation to rethink the very standpoint from which scientific knowledge arises.

But there is an even stronger version of this methodological claim that goes beyond acknowledging the first-person perspective as an inescapable condition of science: it maintains that the
natural sciences should \textit{actively incorporate} this perspective into their own
framework. Merleau-Ponty\index{Merleau-Ponty, Maurice} articulated this demand most explicitly:
\begin{quote}
But a physics that has learned to situate the physicist physically, a psychology that
has learned to situate the psychologist in the socio-historical world, have lost the
illusion of the absolute view from above: they do not only tolerate, they enjoin a
radical examination of our belongingness to the world before all science.
(\citealt[27]{MerleauPonty1968})
\end{quote}

For Merleau-Ponty\index{Merleau-Ponty, Maurice}, quantum mechanics represents the closest approximation to this
new kind of science---one that, unlike classical physics, does not merely
\enquote{posit nature as an object spread out in front of us, [but rather] places its own
object \textit{and its relation to this object in question}} 
\cite[85; our emphasis]{MerleauPonty2003}. In this sense, quantum mechanics may be understood as the science that
investigates the \enquote{relations between the observer and the observed} \cite[15]{MerleauPonty1968}.

The correlation\index{correlation} between subject and object is likewise a central theme in Husserl\index{Husserl, Edmund}’s work. In the \textit{Crisis}, he speaks of a \enquote{universal a priori of correlation\index{correlation} between experienced object and manners of givenness} \cite[166]{Husserl1970}. He underscores the importance of this doctrine by noting that his \enquote{life-work has been dominated by the task of systematically elaborating on this a priori of correlation\index{correlation}} \cite[166, note]{Husserl1970}. The following passage offers further clarification of what Husserl\index{Husserl, Edmund} means by
this correlation\index{correlation}:

\begin{quote}
Object, objective being and consciousness belong a priori and inseparably together; and if we can attribute to each consciousness a consciousness-Ego, an Ego belongs to this essential correlation\index{correlation} as well. A possible Ego and a specifically determined possible consciousness are ascribed to each and every possible objective being. \cite[73]{Husserl2003}
\end{quote}

In related contexts, Husserl\index{Husserl, Edmund} often maintains that subjectivity \textit{constitutes}
objectivity. Yet there is little consensus in the secondary literature on how to interpret his talk of constitution\index{constitution} and correlation\index{correlation}. These notions have arguably never been made fully precise, and it remains unclear how exactly the experiencing subject and the experienced object are correlated---or in what sense subjectivity can be said to constitute objectivity. Interestingly, quantum mechanics may offer a way to render these claims more concrete. The quantum formalism itself can be read as encoding a structured relation between subject and object. In this spirit, Steven French has argued that the phenomenology\index{phenomenology}-inspired interpretation of quantum mechanics developed by \cite{LondonBauer1983} can be seen as \enquote{completing the \textit{Crisis}} \cite[Chapter7]{French2023}. We will come back to this issue in more detail in the next section.

Yet before turning to London\index{London, Fritz} and Bauer in detail, it is worth taking a step back to the
themes of mathematization\index{mathematization} and the nature of space. In the context of WFR, one might be tempted to claim that, from a phenomenological standpoint, the three-dimensional Euclidean space of everyday experience is the \enquote{real} one. Yet for Husserl\index{Husserl, Edmund}, such a claim already involves a problematic reification of mathematical concepts. As he puts it, \enquote{the defined Euclidean manifold behaves to space exactly like the number 2 to any concrete Two, any concrete group of two things} \cite[266]{Husserl1996}. The Euclidean manifold and the number two are abstract mathematical constructs; physical space and concrete pairs of things are not. Mathematical objects may represent aspects of physical reality, but they always do so only approximately, never exhaustively.

Among physicists, perhaps the one who came closest to such a line of thought is John Wheeler\index{Wheeler, John Archibald}. In a characteristically transcendental\index{transcendental} spirit, Wheeler\index{Wheeler, John Archibald} suggested that modern physics compels us to replace the question \enquote{Why does space have three dimensions?} with the 
question \enquote{How does the world manage to give the
impression that it has three dimensions?} \cite[351]{Wheeler1980a}. For Wheeler\index{Wheeler, John Archibald}, as for Husserl\index{Husserl, Edmund}, mathematical space must be sharply distinguished from physical space, and mathematical models and concepts can only ever \textit{approximate} physical reality. As he puts it, \enquote{How else can we look at \enquote{space} and \enquote{dimensionality} except as approximate words for an underpinning, a substrate, a \enquote{pregeometry,} that has no such property as dimension?} \cite[351]{Wheeler1980a}. 

Wheeler\index{Wheeler, John Archibald}’s notion of \enquote{pregeometry} has a distinctly Husserlian resonance and
recurs throughout his later writings. In this context, Wheeler\index{Wheeler, John Archibald} laments that \enquote{[a]ll of these investigations [we typically find in physics] treat spacetime as a pre-existing continuum; they do not look at it as an approximation to an underlying structure, a pregeometry, a substrate of quite a different kind} \cite[3]{Wheeler1980b}. Reflecting on the relationship between mathematics and pregeometry, he writes: \enquote{In the end we are led back from mathematics to physics in the search for a clue to pregeometry. The only thing that could be worse than not finding pregeometry automatically contained in mathematics would be finding it automatically contained in mathematics} \cite[4f.]{Wheeler1980b}. For Wheeler\index{Wheeler, John Archibald}, this is because the foundational structure of physics is not to be uncovered by pure mathematics but by what he calls \enquote{the quantum, the overarching principle of all physics} \cite[5]{Wheeler1980b}. And what is this principle? In his oft-cited formulation: \enquote{No elementary phenomenon is a phenomenon until it is an observed (registered) phenomenon} \cite[5]{Wheeler1980b}. Variations of this claim appear throughout Wheeler\index{Wheeler, John Archibald}’s oeuvre (e.g., \citealt{Wheeler79,Wheeler1980a}) and form the basis of his idea that quantum mechanics reveals a \enquote{participatory universe}, in which \enquote{observer-participancy} is the key to understanding the world. 

It thus appears that Wheeler\index{Wheeler, John Archibald}, much like Merleau-Ponty\index{Merleau-Ponty, Maurice}, takes quantum theory—precisely
as a \textit{fundamental} physical theory---to be concerned not with delivering a
third-person description of an external reality, but with thematizing the very
relationship between subject and object. In this respect, quantum mechanics becomes a
theory that reflexively includes the conditions of reality's capacity to become an object of observation. 

This brings us to the central question of the remainder of this chapter. Could it be that the phenomenological insights into the relationship between the experiencing subject and the experienced object find unexpected support in quantum mechanics? More specifically, might quantum theory itself \textit{encode} a form of correlationalism---one in which the wave function\index{wave function} represents not a detached external reality but the very relation between subject and object? In other words, does quantum mechanics exemplify what phenomenologists should, in principle, expect from science?

At first glance, this may sound like a strange suggestion. The wave function\index{wave function} is, after all, the central object of one of our most fundamental and successful \textit{natural} sciences, whereas phenomenology\index{phenomenology} is a reflective, \textit{a priori} inquiry into the structures of experience. Yet if we take both phenomenology\index{phenomenology} and certain prominent interpretations of
quantum mechanics seriously, the idea that their insights might converge no longer seems so far-fetched.

As is well known—and as we shall see again in the next section—there exists a long tradition in physics that places \textit{experience} at the very centre of quantum mechanics. Perhaps most prominently, Niels Bohr\index{Bohr, Niels} maintained that \enquote{physics is to be regarded not so much as the study of something \textit{a priori} given, but rather as the development of methods for ordering and surveying human experience} \cite[10]{Bohr63}. In a similar vein, Werner Heisenberg\index{Heisenberg, Werner} opened his 1925 paper---one of the founding documents of modern quantum mechanics---with the statement that \enquote{[t]he objective of this work is to lay the foundations for a theory of quantum mechanics based exclusively on relations between quantities that are in principle observable} (Heisenberg, as cited in \citealt[20]{Rovelli2021}).

Erwin Schrödinger\index{Schrödinger, Erwin}, in turn, came to the striking conclusion that \enquote{\textit{Quantum mechanics forbids statements about what really exists---statements about the object. Its statements deal only with the object–subject relation}} (\citealt[490]{Schrödinger2011}; cited in \citealt[753]{FuchsMerminSchack}). Wolfgang Pauli echoed this sentiment when he wrote that in quantum mechanics \enquote{we do not assume any longer the \textit{detached observer}, occurring in the idealizations of this classical type of theory, but an observer who by his indeterminable effects creates a new situation, theoretically described as a new state of the observed system} \cite[33]{Pauli1994}. Hermann Weyl\index{Weyl, Hermann} expressed a similar thought, arguing that modern physics shows that science \enquote{does not state and describe states of affairs---\enquote{Things are so and so}---but that it constructs symbols by means of which it \enquote{represents} the world of appearances} \cite[83]{Weyl2009}, and that if \enquote{we try to untie the real world from the observations, we are left with only a mathematical scheme. Quantum physics necessarily arrives at this decisive insight into the relationship of subject and object} \cite[150]{Weyl2009}.

In the twenty-first century, this line of thought has been taken up by physicists such as Philip Goyal, who argues that the fundamental lesson of quantum mechanics is that science is not meant to provide \enquote{a description of \textit{reality in itself} [but] a description of \textit{reality as experienced by an agent}} \cite[584]{Goyal2012}. Similarly, the proponents of QBism\index{QBism} explicitly affirm their agreement with Bohr\index{Bohr, Niels} \enquote{that the primitive concept of \textit{experience} is fundamental to an understanding of science} \cite[749]{FuchsMerminSchack}.

If this reading is correct—if quantum mechanics, and the wave function\index{wave function} in particular, describe not an objective reality \textit{in itself} but something like the structure of experience or the relation between subject and object—then the parallels with phenomenology\index{phenomenology} are hardly coincidental. On this view, the apparent resonance between phenomenological insights and the lessons of quantum mechanics would not be a mere analogy but a reflection of shared underlying commitments. A deeper grasp of phenomenology\index{phenomenology} might therefore yield a deeper understanding of quantum theory itself.

In what follows, we turn to an interpretation of quantum mechanics that was shaped from the outset by phenomenological ideas: the London–Bauer interpretation. In Section~\ref{sec5}, we will consider what this interpretation suggests about the ontological status of the wave function\index{wave function}.

\section{The London-Bauer interpretation of quantum mechanics}\label{sec4}

As we pointed out earlier, the philosophical nuances behind LBI\index{London-Bauer-Interpretation} (the London-Bauer-Interpretation) cannot be grasped without some basic tenets of Husserlian phenomenology\index{phenomenology}. With this framework now in place, we can finally turn to what LBI\index{London-Bauer-Interpretation} aims to achieve. To set the stage, recall where we left off in the debate about WFR, namely Ney’s account of the metaphilosophical background that shapes her conception of what it means to interpret quantum mechanics. According to her, quantum mechanics is \enquote{intended to be descriptive of an object or objects that exist \textit{independently of us or any other observer}} (\citealt[530]{Ney2012}; our emphasis), that is, it \enquote{describes what the world is like, \textit{independent of us as observers}} (\citealt[3110]{Ney2015}; our emphasis). The task of interpretation is therefore to understand quantum mechanics in a way that brings the theory into line with this general aim.

If one surveys the contemporary landscape in the foundations of quantum mechanics, it is immediately clear how widespread this sentiment is: many who engage in foundational debates take the aim of interpretation to be a description of reality without observers (cf., e.g., \citealt[]{Maudlin2019, Goldstein1998, Bell2004}). Yet, as dominant as this view may be today, we should not forget, as we pointed out in the preceding section, that several of the founding figures of quantum physics held quite different ideas about the lessons to be drawn from the new theory. For instance, in a famous passage, Werner Heisenberg\index{Heisenberg, Werner} writes that

\begin{quote}
The aim is no longer an understanding of atoms ‘in themselves’ […]. From the start
we are involved in the argument between nature and man in which science plays only a
part, so that the common division of the world into subject and object […] is no longer
adequate […]. Thus, even in science, the object of research is no longer nature but man’s
investigation of nature. \citep[58]{Heisenberg1958}
\end{quote}

\noindent A similar point is made by Niels Bohr\index{Bohr, Niels} who openly rejects any \enquote{sharp separation between object and subject} \citep[96]{Bohr1961}, or by Erwin Schrödinger\index{Schrödinger, Erwin} who writes that%

\begin{quote}
Most of us today feel that this necessary abandonment of a purely objective description
of Nature is a profound change in the physical concept of the world. We feel it as a painful limitation of our right to truth and clarity, that our symbols and formulas and the pictures connected with them do not represent an object independent of the observer but only the relation of subject to object. But is this relation not basically the one true reality that we know? (Schrödinger, quoted in \citealt[251]{Moore1989})
\end{quote}

\noindent Our aim here is not to enter into a detailed discussion of the philosophical or historical background of these quotes. Rather, we want to highlight a mindset that shaped the Zeitgeist of the early years of quantum theory and that also forms the backdrop of LBI\index{London-Bauer-Interpretation}. Consider, for example, the following extended passage:

\begin{quote}
The philosophical point of departure of the [classical] theory, the idea of an observable world, totally independent of the observer, was a vacuous idea: Without intending to set up a theory of knowledge, although they were guided by a rather questionable philosophy, physicists were so to speak trapped in spite of themselves into discovering that the formalism of quantum mechanics already implies a well-defined theory of the relation between the object and the observer, a relation quite different from that implicit in naive realism, which had seemed, until then, one of the indispensable foundation stones of every natural science. \citep[220]{LondonBauer1983}
\end{quote}

\noindent This is a rich passage to which we will return repeatedly. For now, however, two aspects deserve particular attention. First, we must clarify what London\index{London, Fritz} and Bauer mean by \enquote{naive realism} and why they regard it as a \enquote{vacuous idea}. Second, we need to examine their claim that quantum mechanics \enquote{already implies a well-defined theory of the relation between the object and the observer}. Since, on our reading, these questions are closely interrelated, it is best to address them together. The most straightforward way to do so is by following London\index{London, Fritz} and Bauer’s own argumentative structure, which begins with von Neumann’s influential distinction between \enquote{Process I\index{Process I}} and \enquote{Process II\index{Process II}}.

One reason why LBI\index{London-Bauer-Interpretation} has at times been misconstrued as no more than an accessible summary of John von Neumann’s landmark \textit{Mathematical Foundations of Quantum Mechanics} (\citeyear{vonNeumann2018}) is that London\index{London, Fritz} and Bauer explicitly frame their discussion of measurement around the now-canonical distinction between two types of processes in quantum theory. The first, which von Neumann calls \textit{Process I\index{Process I}}, is the unitary, deterministic, and reversible evolution of the wave function\index{wave function} according to the Schrödinger equation. If one looks at the practice of physics, this is the aspect of quantum theory that does most of the heavy lifting: it is what allows us to predict interference patterns, energy spectra, scattering cross sections, and so on. Unsurprisingly, realists in the contemporary debate often take \textit{Process I\index{Process I}} to define the “true” dynamics of the system. On this reading, the wave function\index{wave function} represents a physical entity that evolves in time quite independently of any observer, and the task of interpretation is to spell out what kind of object it must be for such an evolution to make sense.

Yet the story does not end here, because von Neumann also distinguishes what he calls \textit{Process II\index{Process II}}. This process is very different in character: it is indeterministic, irreversible, and involves a sudden change in the wave function\index{wave function} whenever a definite outcome is registered. In the textbook presentation this is written into the so-called projection postulate, where the superposition\index{superposition} of possible outcomes is replaced by the actual result of a measurement. It is here that the notorious \enquote{collapse\index{collapse} of the wave function\index{wave function}} enters the picture.

As we have indicated earlier, the puzzle, then, is to understand what happens in the transition from Process I\index{Process I} to Process II\index{Process II}. If the wave function\index{wave function} is taken to represent a genuine physical entity—whether conceived as a field on configuration space or as a state vector in Hilbert space—then the question arises: how can it suddenly undergo a discontinuous change that is neither predicted by the Schrödinger equation nor reversible in principle? Should we think of the collapse\index{collapse} as a physical event triggered by an interaction with a measurement apparatus? Should we instead treat it as a mere bookkeeping device, reflecting only the limits of our knowledge? Or is it something else entirely? These are the questions that have animated generations of discussions about the measurement problem\index{measurement problem}. 

Before we look at how London\index{London, Fritz} and Bauer propose to deal with the issue at hand, it is helpful to take a step back and consider more carefully what \textit{Process I\index{Process I}} amounts to. When we speak of Process I\index{Process I}, we are not dealing with a bare formalism in the void: a specific experimental context is presupposed.

First, an observable is chosen. In practice this means fixing which physical quantity is to be measured—say, the spin of an electron along the $z$-axis or the position of a photon on a screen. This choice determines the relevant basis in which outcomes will be registered. Second, the system is prepared in an initial state, often (but not always) a superposition\index{superposition} relative to that basis. For example,
\[
\lvert \psi \rangle = \alpha \lvert \uparrow_z \rangle + \beta \lvert \downarrow_z \rangle,
\]
with $|\alpha|^2 + |\beta|^2 = 1$.

Third, the measurement interaction begins: the system couples to the apparatus. In a Stern–Gerlach experiment, for instance, the electron enters a magnetic field; in a double-slit experiment, the photon wave packet reaches the detector screen. The mathematics of quantum theory describes this as a unitary evolution that correlates system states with distinct possible pointer states of the apparatus.

The result is an entangled state of system and apparatus, schematically
\[
\lvert \Psi \rangle = \alpha \lvert \uparrow_z \rangle \otimes \lvert \text{pointer up} \rangle \;+\; 
\beta \lvert \downarrow_z \rangle \otimes \lvert \text{pointer down} \rangle.
\]
At this stage, the evolution is fully deterministic and reversible: the initial superposition\index{superposition} of the system has simply spread into an entangled superposition in which the system and apparatus are correlated. Importantly, nothing in this description commits us to a specific ontology: the same mathematics can be read in standard Hilbert-space language or, if one prefers, in terms of a field evolving in configuration space.

What is crucial from the perspective of LBI\index{London-Bauer-Interpretation} is that, so described, Process I\index{Process I} is not yet a measurement in the full sense of the term. The unitary evolution delivers correlations between system states and possible pointer positions. Yet at this stage the apparatus is still described as being in a superposed relation to different outcomes.  

In the spin example, the apparatus ends up correlated with both possibilities, schematically
\[
\alpha \lvert \uparrow_z \rangle \otimes \lvert \text{pointer up} \rangle \;+\; 
\beta \lvert \downarrow_z \rangle \otimes \lvert \text{pointer down} \rangle.
\]
This is not to say that the description is “merely mathematical.” Rather, it shows that the formalism itself stops short of telling us which of the possible outcomes has been realized. In other words, Process I\index{Process I} provides a \textit{horizon\index{horizon} of possibilities}: the structure of potential outcomes is fully spelled out, but no single result has yet been constituted as actual.\footnote{It is important not to misinterpret this limitation of Process I\index{Process I} as if the formalism were “merely mathematical” in contrast to some separate physical process. The unitary dynamics already encodes the concrete couplings between system and apparatus (magnetic gradients in a Stern–Gerlach setup, photon–detector interactions, etc.), and it makes experimentally testable predictions. The resulting entangled state captures real physical correlations. What Process I\index{Process I} does \emph{not} provide is the singling out of one of these correlations as the \emph{actual} outcome. That further step—how objectivity itself is constituted—is precisely what London\index{London, Fritz} and Bauer want to address, as we shall see shortly.} This is why, according to LBI\index{London-Bauer-Interpretation}, Process I\index{Process I} alone does not capture what measurement really is, why \enquote{a coupling, even with a measuring device, is not yet a measurement} \citep[251]{LondonBauer1983}. The pressing question, then, is: what is missing?

At this juncture, a familiar realist manoeuvre suggests itself: insist that the outcome was there all along, and that the interaction between quantum system and apparatus merely reveals it. But if we extend London\index{London, Fritz} and Bauer’s line of reasoning to this strategy, it becomes clear why naïve realism, for them, is nothing but a “vacuous idea”. The vacuity lies, first of all, in the fact that nothing in the mathematics of quantum mechanics points to a single, pre-existing result. Process I\index{Process I} generates an entangled superposition\index{superposition} that encodes correlations between system states and apparatus states, but it does not privilege one of these correlations as the actual outcome. To claim otherwise is simply to add determinacy by fiat.  

Second, even if one were to postulate that the outcome was always already fixed, this does not address the real problem. Such a move merely restates the measurement problem\index{measurement problem} rather than solving it, because it leaves unexplained how one particular result becomes available as the outcome of the measurement. In other words, the realist manoeuvre dodges the explanatory task that quantum mechanics itself has put before us: the task of accounting for the transition from a horizon\index{horizon} of possibilities to a constituted result.

 But what, then, is London\index{London, Fritz} and Bauer’s proposed solution? To begin with, LBI\index{London-Bauer-Interpretation}'s strategy is to widen the scope of the description. Instead of stopping at the coupled system–apparatus pair, they insist that we must also include the \textit{observer}. Formally speaking, this means that the global state after the interaction is not just a superposition\index{superposition} of correlations between system and apparatus, but an entangled state that also encompasses the observer:
\[
\lvert \Psi \rangle = \sum_i c_i\, \lvert s_i \rangle \otimes \lvert A_i \rangle \otimes \lvert O_i \rangle .
\]
At this level of description, nothing has changed in the structure: the state is still a superposition\index{superposition} of correlations, only now extended to include the observer as well.  

What makes a decisive difference, however, is that the observer occupies a distinctive position within this triad. Formally speaking, system, apparatus, and observer are on equal footing: each is a subsystem of the same entangled state. Yet London\index{London, Fritz} and Bauer insist that the observer is set apart in one crucial respect. Whereas the system and the apparatus are simply carried along by the unitary dynamics, the observer \enquote{has \textit{with himself} relations of a very special character, [namely] a characteristic and quite familiar faculty which we can call the \enquote{faculty of introspection}} \citep[252]{LondonBauer1983}. From a phenomenological perspective, it is clearer to speak here of \textit{reflection}, and to describe the product of such reflection as a form of \textit{immanent knowledge}.\footnote{We cannot enter this debate in detail here, but the reasons why in phenomenology\index{phenomenology} \enquote{reflection} is preferred over \enquote{introspection} are that the latter misleadingly suggests an inner observation of psychic states, whereas phenomenological reflection denotes a shift of stance: from simply living through an experience to reflecting on it, from straightforwardly engaging with the natural environment to a more detached focus on the experience of being so engaged. Cf., for further discussion, \citep[]{Cerbone2012, Thomasson2005}.}  

This capacity for reflection is crucial because it is what allows the observer to “step back” and take a stance toward her own state within the entangled whole. Only she, and not the system or the apparatus, can say of herself: “I am in state $w_k$.” The self-ascription “$w_k$” is, in London\index{London, Fritz} and Bauer’s terminology, the condition that grounds the constitution\index{constitution} of a definite fact, expressed as “$F = f_k$.” According to LBI\index{London-Bauer-Interpretation}, it is precisely this reflective act and this capacity to pass a judgment on one’s own state that grounds the shift from a mere horizon\index{horizon} of possibilities to a constituted objectivity.

The moment the observer declares “$F = f_k$”, \enquote{the observer established his own framework of objectivity and acquires a new piece of information about the object in question} \citep[252]{LondonBauer1983}. What had been an open superposition\index{superposition} of correlations is resolved into an objectivity that holds not just for the observer herself but for the intersubjective world of science. collapse\index{collapse}, on this view, is not an unexplained physical jolt injected into the dynamics from the outside, nor a brute stipulation layered onto the mathematics. It is the name we give to the constitutive act by which reflection transforms a spread of possibilities into an objective fact. In other words, what the realist treats as a puzzling discontinuity in the reality described by the formalism is, for London\index{London, Fritz} and Bauer, the moment in which objectivity itself is brought forth. In their own words:

\begin{quote}
Thus it is not a mysterious interaction between the apparatus and the object that produces a new $\psi$ for the system during measurement. It is only the consciousness of an \enquote{I} who can separate himself from the former function $\psi$(\textit{x, y, z}) and, by virtue of his observation, \textit{set up a new objectivity} in attributing to the object henceforward a new function “$\psi(x) = u_k(x)$”. \citep[252]{LondonBauer1983}
\end{quote}

\noindent At this point, one might worry that LBI\index{London-Bauer-Interpretation} risks sliding into two well-known pitfalls. The first is the charge of \textit{idealism}: doesn’t the talk of reflection constituting objective facts amount to saying that observers \textit{create} reality simply by looking? The second is the suspicion that what we are left with is just the notorious “consciousness causes collapse\index{collapse}” story, according to which a trans-physical consciousness mysteriously intervenes in the physical world to trigger the collapse\index{collapse} of $\psi$. If that were the substance of LBI\index{London-Bauer-Interpretation}, it would fall prey to the classic criticisms of Putnam and Shimony, who argued that such dualist collapse\index{collapse} stories are both conceptually incoherent and physically unmotivated \citep[]{Shimony1963, Putnam1979}.

Let us take the second worry first. London\index{London, Fritz} and Bauer make it clear that their account does not tamper with the physics. The Hamiltonian remains the same, the Schrödinger equation still governs the dynamics, and the unitary evolution of Process I\index{Process I} unfolds exactly as before. Nothing in their proposal adds a new physical ingredient or postulates an extra causal mechanism. What changes is not the physical dynamics but the position of the observer within it. Whereas in Process I\index{Process I} the observer is formally just another subsystem entangled along with system and apparatus, she also has the distinctive capacity to reflect on her own state. By stepping back from the entangled flow and declaring, “I am in state $w_k$,” she constitutes the situation as one in which a definite outcome has been obtained.  

Collapse\index{collapse}, therefore, is not conceived as a physical jolt imposed from the outside, nor as the intervention of a mind beyond nature. For precisely this reason, London\index{London, Fritz} and Bauer’s proposal should not be confused with the “consciousness causes collapse\index{collapse}” line that later became associated with Eugene Wigner.\footnote{Unfortunately, we cannot enter this discussion in detail here. It is worth noting, however, that even in Wigner’s case the situation is more nuanced than the familiar caricature suggests. Although his name has become shorthand for a “consciousness causes collapse\index{collapse}” doctrine, Wigner’s own views were more ambivalent and shifted over time, and there is considerable debate in the secondary literature about how literally his remarks on the role of consciousness should be taken. See, for further details, \citep[chapter 3]{French2023}.} Nothing in LBI\index{London-Bauer-Interpretation} suggests that consciousness reaches into the physical system to force a discontinuity in its dynamics. Instead, collapse\index{collapse} is just the name we give to the appearance of definiteness from within: the transition whereby an open horizon\index{horizon} of entangled possibilities is articulated into a determinate fact by an act of reflection.

To make this clearer, and also to respond to the first worry regarding the charge of idealism, compare the quantum case with the structure of simple perception. While we are immersed in the flow of lived experience, the perceptual encounter with our surroundings is a constant dynamic probing of a structured, yet open horizon\index{horizon} of possibilities. Think, for example, of the ordinary activity of driving a car. If you are experienced, you do not constantly monitor your epistemic position with respect to the steering wheel or the gear stick. Your body and attention are oriented toward the situation as a whole: the horizon\index{horizon} of what might happen in traffic, the readiness to turn, brake, accelerate. It is typically only when something goes wrong—a sudden noise, a wrong gear, a swerve—that you take a reflective step back, freeze the episode, and explicitly notice: “I am here, now, in this epistemic position.”  

London\index{London, Fritz} and Bauer’s point is that the quantum case works in an analogous way. Process I\index{Process I} corresponds to a horizon\index{horizon} of correlated possibilities: the superposed, entangled state of system, apparatus, and observer. It is only when a measurement occurs that the observer takes a reflective step back, becomes aware of her position within the entangled whole, and judges: “I am in state $w_k$.” In this act, the horizon\index{horizon} of possibilities is articulated into a determinate result. London\index{London, Fritz} and Bauer describe this moment as the constitution\index{constitution} of a new objectivity, expressed as “$F = f_k$.” collapse\index{collapse}, in their account, is not a physical process over and above the unitary dynamics, but a shift of stance: the separation of the entangled correlation\index{correlation} into an Ego-pole and an Object-pole. What appears to the realist as a mysterious discontinuity is, on this view, the ordinary structure of how objectivity is constituted.

Now, to restate the question from before: is this idealism? In one sense, it is—if idealism is taken to mean the rejection of the idea that reality is simply a brute, uninterpreted \enquote{in-itself} wholly independent of the structures through which it is encountered. According to LBI\index{London-Bauer-Interpretation}, as for phenomenology\index{phenomenology} more broadly, reality is always mediated by a process of meaning-bestowal: what counts as an object, what shows up as determinate, is constituted out of a correlation\index{correlation} between subject and world. In this sense, subject and reality are never strictly separate; they are already intertwined within a horizon\index{horizon} of possibilities from which definite outcomes are constituted. But this should not be confused with the caricature of the subject conjuring outcomes at will. The subject cannot decide to see a green light turn red or a gear stick turn into a steering wheel. Likewise, the quantum observer cannot simply create outcomes ex nihilo. What the reflective act accomplishes is the articulation of a determinate result within the structured horizon\index{horizon} already set up by the dynamics. Seen in this light, there is less difference than one might think between the constitution\index{constitution} of objectivity in ordinary perceptual experience and in quantum measurement. The novelty is that in quantum mechanics this correlationist structure is not merely presupposed in the background of lived experience but is explicitly encoded in the very formalism of $\psi$. We take it that this is the ultimate lesson to be drawn from LBI\index{London-Bauer-Interpretation}.

\section{Wave function realism is realism about... what?}\label{sec5}

To understand what LBI\index{London-Bauer-Interpretation} is claiming, it is enough to start with a simple idea: the wave function does not describe a hidden physical thing, but the structure of the possibilities that are available before a definite outcome is registered. If the measurement transition is, as we have argued, not a physical jump but the moment at which one of these possibilities becomes the actual outcome for an observer, then the wave function captures the organized field of such possibilities. In phenomenological terms, one can say that $\psi$ gives a \textit{formal expression of the horizonal structure of how a world can show up for us}. We take this to be the ultimate take-home message of LBI\index{London-Bauer-Interpretation}. From this view, then, the wave function\index{wave function} does not describe a hidden physical entity underlying appearances, but rather encodes the correlational fabric within which appearances acquire determinacy in the first place. The wave function\index{wave function} represents, as it were, the open field of possible correlations---an abstract analogue to what phenomenology\index{phenomenology} calls the \textit{horizon\index{horizon} of experience}.

It is important to head off a familiar misunderstanding that arises when the idea of a horizon is first introduced. Everyday illustrations---such as the cup that shows only one of its sides while the rest is anticipated---can make it seem as if the horizon were a personal backdrop, tied to the particular perceptions of an individual subject. But in Husserl’s usage, \enquote{horizon} is not a psychological notion at all. It is a structural concept: it refers to the general conditions under which anything can appear as part of a world in the first place. In that sense, there is no \enquote{my} horizon or \enquote{your} horizon in the empirical sense; rather, there is a shared structural openness within experience that allows objects to show themselves as more than what is presently given. Horizon, so understood, is \textit{impersonal}. It does not describe the private field of a subject but the general way in which experience remains open to further determination.

Read in this light, the formalism of quantum mechanics can be seen as giving a mathematical articulation of this very structure. The superposition\index{superposition} described by $\psi$---that is, the coexistence of multiple possible outcomes within a single coherent state---mirrors the horizonal openness of experience before reflective constitution\index{constitution}. The act of measurement, as London\index{London, Fritz} and Bauer interpret it, corresponds to the reflective articulation of one possible correlation\index{correlation} as actual. In phenomenological language, it marks the passage from an indeterminate horizon\index{horizon} of potential givenness to a constituted objectivity. The mathematics of quantum theory, then, is not the representation of a mind-independent substrate, but the formal codification of the conditions under which a world of determinate objects can appear at all.

There is another potential misunderstanding that should be addressed at this point. We are not the first to suggest that the key to understanding quantum mechanics phenomenologically may lie in treating the wave function\index{wave function} as a mathematical representation of the horizonal structure of experience. In her attempt to interpret \enquote{QBism\index{QBism} [...] as a phenomenological reading of quantum mechanics}, Laura de La Tremblaye advances a proposal that, at first sight, appears strikingly similar to our own:

\begin{quote}
[T]he QBist [modifies] her state vector according to the results [...] obtained, in order to optimize his [sic!] subsequent bets. The same method is found in Husserlian phenomenology\index{phenomenology}. Several futures were open to me at the time of the fall of my cup. A horizon\index{horizon} of still undetermined possibilities was present before the cup hit the ground. Among the endless possibilities prefigured by the horizon\index{horizon}, there were two scenarios: the scenario in which the cup breaks, and the scenario in which the cup survives the fall. After the cup burst into pieces, the anticipation of the cup remaining undamaged is frustrated, and other anticipations come to the fore. A new horizon\index{horizon} is opening up, just as undetermined as the previous one, but redefined. [...] The cup is the analogue of the microsystem, the perceptual horizon\index{horizon} parallels the QBist quantum state, the perceptual act corresponds to the physicist's measurement and the modification of my possible horizon corresponds to the modification of the state vector after measurement. \citep[255]{Tremblaye2020}
\end{quote}

Steven French has reacted critically to this suggestion in his groundbreaking book on LBI\index{London-Bauer-Interpretation}. In short, he raises two main objections against de La Tremblaye's proposal. First, he questions \enquote{whether we can straightforwardly draw parallels between our everyday experiences, embedded as they are in the \enquote{life-world\index{life-world}} and those that arise in the form of what the QBist calls \enquote{kicks} from the world manifested in the spin measurement} \citep[197]{French2023}. The background of this concern is easy to see. Throughout her exposition of the phenomenological concept of \enquote{horizon}, de La Tremblaye relies almost exclusively on examples from perceptual experience, defining the horizon\index{horizon} explicitly as the \enquote{\textit{perceptual} horizon\index{horizon} [...] constituting the \enquote{\textit{perceptual} process}} (\citealt[253]{Tremblaye2020}; our italics). Given this tendency to equate the horizon\index{horizon} with sensory---and predominantly visual---experience, French's worry is natural: why should a structural feature of the way we experience, say, coffee cups or falling objects have any relevance for phenomena as far removed from life-world\index{life-world} experience as quantum systems?

Yet French’s worry, though understandable, ultimately rests on a misconception of what the term horizon\index{horizon} denotes in the phenomenological tradition. As already emphasized, horizonality is not a feature peculiar to perceptual experience---let alone to the visual contemplation of coffee cups or falling objects---but a formal structure of world-manifestation as such. In Husserl’s sense, every mode of givenness---whether sensory, affective, memorial, imaginative, or theoretical---unfolds within a horizon\index{horizon} that both delimits and opens the field of possible further experience. What differs is not the existence of a horizon\index{horizon}, but the manner in which it manifests. In perception, the horizon\index{horizon} appears as the co-givenness of unseen aspects of the thing; in auditory experience, as the anticipation of how a melody will continue; in the grasp of an abstract object, as the network of inferential or conceptual possibilities that radiate from an evident insight.\footnote{\textit{Insight} here is used in the phenomenological sense of an act that presents something as evident---not in the sense of an immediate or mystical apprehension of mathematical objects. Even in mathematics, proofs typically culminate in such acts of evident grasp: the structure established by the proof becomes \enquote{seen} as holding.} phenomenology\index{phenomenology}’s task, then, is not merely to catalogue these variations but to reveal in them the invariant structure of horizonal givenness itself—the formal condition under which any world, regardless of modality, can appear as a world at all.

From this standpoint, there is nothing arbitrary in extending the notion of horizon\index{horizon} to quantum phenomena. If quantum mechanics is, as London\index{London, Fritz} and Bauer suggest, a formal codification of the correlation\index{correlation} between observer and observed, then the quantum horizon\index{horizon} is simply the structured openness of possible correlations that precede the constitution\index{constitution} of a determinate outcome. The wave function\index{wave function} mathematically articulates this impersonal field of possibilities in much the same way that the horizonal structure of experience articulates the space of possible appearances. The point is thus not to assimilate quantum mechanics to the sensory world of everyday objects, but to recognize in both cases a shared transcendental\index{transcendental} structure: the world---any world---shows up only within a horizon\index{horizon} of indeterminate possibilities that can be reflectively articulated into determinate objectivities.

French’s second concern is that, as he puts it, \enquote{what is portrayed as the basis of QBism\index{QBism}---namely the first-person perspective---must be modified [...] in order to maintain its alignment with [the phenomenological notion of horizon]} \citep[200]{French2023}. The background to this remark becomes clear once we recall the passage from de La Tremblaye discussed earlier. For it is no accident that she speaks of the state vector rather than the wave function\index{wave function}.\footnote{For clarity: by \enquote{state vector} de La Tremblaye is referring to the usual Hilbert-space representative of a quantum state—that is, a vector in a ray. A wave function is simply the coordinate expression of such a vector in the position basis. QBist usage treats whichever representative is chosen as encoding an agent’s personal expectations. Nothing in our discussion turns on the distinction between a ray and one of its representatives, nor on whether one works with the abstract vector or its position-space expression.} This is because, within QBism\index{QBism}, the state vector is understood to encode an agent’s personal expectations about possible measurement outcomes---it is not, even in a refined philosophical sense, usually seen as a representation of the world itself. Her analogy between the horizon\index{horizon} and the quantum formalism therefore operates on the epistemic level of an individual subject’s field of anticipation. And this, as French rightly observes, is precisely where the phenomenological analogy breaks down: the horizon\index{horizon} in Husserl’s sense is not an empirical or psychological feature of a particular consciousness, but a transcendental\index{transcendental} structure that makes possible any experience or world-manifestation whatsoever. To equate it with an agent’s private web of expectations is therefore to miss the very point of the phenomenological notion.

However, as we have already emphasized, our own account of the horizon\index{horizon} differs fundamentally from the conception that motivates French’s criticism. For us, the horizon\index{horizon} is not the experiential backdrop of an individual consciousness but the impersonal and transcendental\index{transcendental} structure that makes any manifestation of a world possible in the first place. As we have explained, it is not something one \enquote{has,} like a private perceptual field, but something within which both subject and object come to be distinguished at all. From this standpoint, French’s objection---though maybe pertinent in assessing de La Tremblaye’s rendering---does not apply to our proposal. LBI\index{London-Bauer-Interpretation}, as we understand it, operates precisely at a thoroughly transcendental\index{transcendental} level: the wave function\index{wave function} does not encode an agent’s personal expectations but articulates the impersonal network of correlations through which objectivity itself can arise.

If we adopt the perspective described here, the question of realism takes on a new form. Are London\index{London, Fritz} and Bauer realists about the wave function\index{wave function}? In one sense, clearly not. If \enquote{realism} means, as it does for Ney and many contemporary authors, the conviction that quantum mechanics aims to describe a reality \textit{minus observers}, then London\index{London, Fritz} and Bauer reject realism outright. For them, the very distinction between observer and observed is not primitive but constituted: reality is never \enquote{there} prior to or independently of the correlational structure that makes it manifest. Yet in another, and in our view deeper, sense they are realists. The wave function\index{wave function}, for London\index{London, Fritz} and Bauer, represents something that is not merely subjective or arbitrary. What it expresses is the transcendental\index{transcendental} structure of world-manifestation itself---the impersonal correlation\index{correlation} through which anything like an objectivity becomes possible. In this respect, their realism is not \textit{objectivist} but \textit{correlational}: $\psi$ stands for the formal structure of the world’s self-presentation.\footnote{Regarding its ontological status, French argues that the wave function\index{wave function} \enquote{can be situated between the $\psi$-epistemic \& $\psi$-ontic accounts} such that in one sense it is both but in the phenomenological sense it is neither \enquote{as it expresses the \textit{fundamental correlative relationship} that sits at the heart of the phenomenological stance} \cite[12]{French2024}; see also \cite[37]{French2025}. The \enquote{collapse} is then conceptualized as a shift from an external to an internal perspective \cite[155]{French2023}. For a critical discussion see \cite{Pienaar2025}, who argues that this position is in danger of leading to \enquote{a many-worlds interpretation with a phenomenological gloss}. Here we note that on our view the \enquote{collapse\index{collapse}} is not so much a shift from an external to an internal perspective but rather one from co-givenness to originary givenness.}

To call this \enquote{realism} may sound paradoxical, but it need not be. LBI\index{London-Bauer-Interpretation} can be seen as advocating a form of \textit{transcendental\index{transcendental} realism}: a recognition that there is something genuinely real---namely, the horizonal structure of givenness---that precedes and conditions the opposition between subject and object. This is not the realism of the laboratory table but the realism of appearance itself. In contrast to the objectivist realism of Ney, which seeks to describe reality as it would be \textit{without} observers, London\index{London, Fritz} and Bauer propose that the very possibility of such a description presupposes the horizonal structure that their interpretation brings to light.

In this sense, French is partly\footnote{As one of us has recently argued, French’s claim that LBI\index{London-Bauer-Interpretation} completes Husserl\index{Husserl, Edmund}’s project in the \textit{Crisis} must be treated with some caution. While French is right that, \textit{epistemologically}, LBI\index{London-Bauer-Interpretation} represents a decisive move toward a phenomenological framework, he underestimates the \enquote{constitutional history} of the mathematical idealities that underlie quantum mechanics. For further discussion, see \citealt{IslamiWiltsche2025}.} correct that London\index{London, Fritz} and Bauer’s reading of quantum mechanics can be seen to complete Husserl\index{Husserl, Edmund}’s \textit{Crisis}. Husserl\index{Husserl, Edmund} had diagnosed the \enquote{forgetting of the life-world\index{life-world}} as the point at which scientific mathematization\index{mathematization} loses sight of its own experiential ground. London\index{London, Fritz} and Bauer, working within the same intellectual milieu, re-inscribe this ground into the formalism of physics itself. The wave function\index{wave function} does not lie behind appearances but articulates their very structure of manifestation. To understand $\psi$ phenomenologically is therefore to recover the link between the empirical sciences and the transcendental\index{transcendental} conditions of their possibility---a link that Husserl\index{Husserl, Edmund} saw slipping away with Galileo and that London\index{London, Fritz} and Bauer, perhaps unwittingly, sought to restore.

From this vantage point, LBI\index{London-Bauer-Interpretation} is realist not about a hidden ontology behind experience but about the structural conditions that make any experience of a world possible. It offers a realism of manifestation rather than of objects, a realism of correlation\index{correlation} rather than of substance. And this, in turn, reconfigures the question with which we began. If the task of interpretation is to say what quantum mechanics tells us about reality, then the answer proposed by London\index{London, Fritz} and Bauer is neither that it reveals a world of entities independent of observers nor that it confines us to the private sphere of subjective belief. Instead, it tells us that the mathematical formalism of quantum theory encodes the very correlation\index{correlation} between observer and world---the dynamic, horizonal field within which both subject and object are constituted.\footnote{Here---as everywhere else in this chapter---\enquote{correlation} is not meant in the statistical sense of a pairing between two sets of empirical states (e.g., time-series data of a mind and external objects). In the phenomenological literature---and in LBI---the correlation between subject and world is a \textit{transcendental structure}: it names the conditions under which subjects and objects can appear as such at all. It is therefore not inferred from observational data, but articulated through a reflective analysis of the very possibility of experience and measurement. Our point is that LBI reads the quantum formalism as encoding this structural relation, not as expressing correlations between empirical sequences of mental and physical states.} In this sense, the wave function\index{wave function} does not represent a thing in the world; it represents the world’s own capacity to appear.\footnote{A similar and perhaps complementary interpretation is offered in \cite{Berghofer2025}. This approach agrees with QBism\index{QBism} that the wave function\index{wave function} is an assignment that encodes the experiential input to the quantum machinery (Born rule) but interprets quantum probabilities as an \textit{objective output} specified in terms of \textit{objective degrees of epistemic justification}, quantifying what the experiencing subject should expect to experience next.}

We are, of course, aware that the account of LBI\index{London-Bauer-Interpretation} offered here can only be a further step, the first ones having been taken in the work of Steven French. To turn it from a historical curiosity into a serious interpretational framework, much work remains to be done on both the phenomenological and the physical sides. Phenomenologically, the relation between reflection, constitution\index{constitution}, and horizonal structure requires further articulation---especially if the act of measurement is to be understood as a reflective transition from potentiality to objectivity without sliding into idealism. On the side of physics, one must examine how LBI\index{London-Bauer-Interpretation} addresses some of the core puzzles that define the contemporary debate: How does a horizon-based account handle Wigner’s friend and its extended versions, where nested observers threaten to collapse the distinction between reflection and interaction? What becomes of locality and separability, given that LBI\index{London-Bauer-Interpretation} relocates objectivity from three-dimensional space to the correlational field of experience? And how are we to understand violations of Bell inequalities within such a framework---are they to be seen as signatures of the non-separability of the transcendental\index{transcendental} correlation\index{correlation} itself? These are large questions, but the direction of travel is clear. If the direction of LBI\index{London-Bauer-Interpretation} is correct, then the challenge is not to reinsert the observer into a pre-given world, but to spell out how the world itself, as mathematically captured by the wave function\index{wave function}, emerges through the horizonal dynamics of correlation.

\backmatter

\bmhead{Acknowledgements}

Earlier versions of this paper were presented at workshops at Università Roma Tre, Italy, and UC Louvain, Belgium. We thank the organizers and respective audiences for valuable feedback. Special thanks go to Steven French, Mauro Dorato, Alyssa Ney, Daniele Pizzocaro, Jordan François, Lucrezia Ravera, and Yanhao He.

\bibliography{references}
\end{document}